# Horizontal SCA Attacks on Binary *kP* Algorithms using Chevallier-Mames Atomic Blocks

Gerald Isheanesu Matungamire[1,3], Alkistis Aikaterini Sigourou[1], Gerrit Schrock[1,2], Zoya Dyka[1,3], Peter Langendoerfer[3] and Ievgen Kabin[1]

[1] *IHP – Leibniz Institute for High Performance Microelectronics,* Frankfurt (Oder), Germany
[2] *Nordhausen University of Applied Sciences*, Nordhausen, Germany
[3] *BTU Cottbus-Senftenberg,* Cottbus, Germany
{matungamire, sigourou, schrock, dyka, langendoerfer, kabin}@ihp-microelectronics.com

*Abstract*— Scalar multiplication *kP* is the operation most frequently targeted in Elliptic Curve (EC) cryptosystems. To protect against single-trace Side-Channel Analysis (SCA) attacks, the atomicity principle and various atomic block patterns have been proposed in the past. In this work we use our software and hardware implementations to demonstrate that binary right-to-left and left-to-right *kP* algorithms, when implemented with Chevallier-Mames atomic block patterns, are still vulnerable to single-trace SCA attacks. The vulnerability remains true for the left-to-right *kP* algorithm with projective coordinate randomization.

## I. INTRODUCTION

Elliptic curves (EC) are the basis of cryptographic protocols for the exchange of secret keys, authentication and digital signature approaches. The most time and energy consuming operation in EC cryptosystems is the EC point-scalar multiplication, denoted as *kP* operation (where *k* is a scalar and *P* is a point on an EC). This operation is mostly attacked with the goal of revealing the value of the secret scalar *k*. Side-channel analysis (SCA) attacks exploit side-channel effects that can be measured during the targeted *kP* execution, for example, current drawn from the power supply or electromagnetic (EM) radiation emitted by the attacked chip. Analysing the power or the EM trace, attackers can reveal the scalar *k*. The resistance of *kP* algorithms to single-trace attacks remains an issue despite many countermeasures proposed in the past. The atomicity principle is a countermeasure strategy based on the representation of *kP* operations as a series of small blocks of mathematical operations. The notation "atomic block" was introduced in [1], proposing decomposing the *kP* operation into many small atomic blocks: each block performs the same sequence of operations causing similar energy consumption over the same execution time, i.e. the shapes of the atomic blocks are very similar to each other. Attackers do not know which block refers to the processing of '0' bit value in the binary representation of *k*, and which to the '1'. A similar idea was proposed in [2], before [1]. In [3], [4], [5], the vulnerability of *kP* algorithms implemented using different atomic block patterns to single-trace SCA attacks was described. The SCA leakages were caused by the data value processed (data-bit) [3] or data flow (address-bit) [4] or by their combination [5]. The last two vulnerabilities were observed for hardware implementations of the atomic patterns [6] and [7]. In this paper, we demonstrate that not only a hardware implementation of binary *kP* algorithms using Chevallier-Mames' atomic patterns [1] can be vulnerable to simple SCA, but also software implementation for microcontrollers, even when implemented using an open source crypto library with time-constant field operations.

## II. IMPLEMENTATION DETAILS AND MEASUREMENTS

We implemented two binary *kP* double-and-add algorithms: the left-to-right and the right-to-left algorithms using atomic block patterns for EC point doubling (further denoted as PD, or 2***Q***) and EC point addition (further denoted as PA, or ***Q***+***P***) operations as proposed in [1]. Each atomic block consists of: multiplication, addition, negation, and addition, known as the MANA atomic block. The atomic patterns for a PD and a PA consist of 10 and 16 MANA atomic blocks, respectively. The sequences of field operations in our hardware and software implementations are given in [8]. For our software implementation, we selected a TI LAUNCHXL-F28379D LaunchPad board [9] with the TMS320F28379D C2000 32-bit dual-core microcontroller as the target device and the open-source cryptographic library FLECC in C [10] due to the availability of the constant-time modular arithmetic functions. The implementation was done for the NIST EC P-256 [11]. For the left-to-right *kP* algorithms, we implemented two versions: without any additional countermeasures (i.e., only atomic blocks were the countermeasure) and a version with randomised projective coordinates of the point ***P*** corresponding to [12]. We did not apply the randomisation of the point coordinates in the right-to-left *kP* algorithm. In our hardware implementation, we realised only the left-to-right *kP* algorithm using the previous IHP design [5] as a basis and changing the data flow corresponding to [1]. All operations were implemented serially and described in VHDL using AMD Vivado Design Suite v2024.2.1 (64-bit). The design was ported for a Digilent Arty Z7-20 board [13] with a Zynq XC7Z020-1CLG400C FPGA.

Measurements of EM emanation during a *kP* operation were performed using a near-field microprobe MFA-R 0.2-75 [14] manufactured by Langer. An integrated circuit scanner Langer ICS 105 [15] was used to guarantee a precise position and orientation of the probe over the board. We measured EM traces on the power supply capacitor C49 located on the back side of the attacked board, see Fig. 1. The EM trace of a *kP* execution on the attacked FPGA was measured also on a power supply capacitor on the back side of the attacked board. The placement and orientation of the EM probe are shown in Fig. 2. All algorithms were executed using the affine coordinate of the base point *G* of the EC P-256 as inputs. We used a 256-bit-long scalar *k* for FPGA execution andonly the 5-bit-long binary scalar $k=11111_2$ for all executions on the microcontroller. We used a Teledyne LeCroy WavePro 604HD oscilloscope to capture the EM trace. The FPGA design runs at 10 MHz, and the trace was measured with the oscilloscope's sampling rate of 1 GS/s. The *kP* execution on the microcontroller (in RAM) was clocked at 100 MHz and measured with 2.5 GS/s.

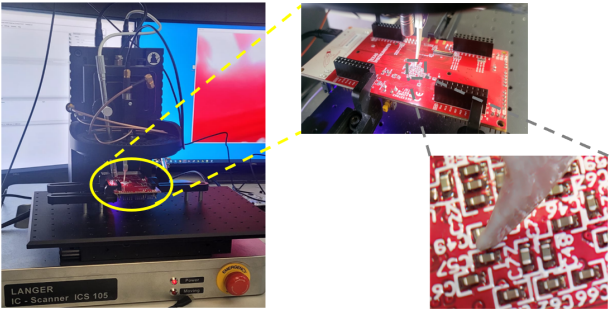

Fig. 1. Measurements of EM trace on the back side of the microcontroller's board: position and orientation of the EM probe are shown, zoomed in.

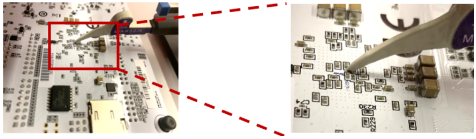

Fig. 2. Measurements of EM trace on the back side of the attacked FPGA board: position and orientation of the EM probe are shown, zoomed in.

### III. ANALYSIS

#### A. Vulnerability to simple SCA

Fig. 3 shows a part of the oscilloscope's waveform for FPGA measurement. The shown part corresponds to the execution of a PA and a PD. Each PA consists of 16 atomic blocks Δ1, …, Δ16; each PD consists of 10 atomic blocks Δ1, … , Δ10. To avoid the successful simple SCA attack, the shapes of all atomic blocks have to be very similar, but it is clearly visible that the shapes of the atomic blocks Δ1-Δ4 and Δ9 in the *Q*+*P* differ significantly from those of the other atomic blocks.

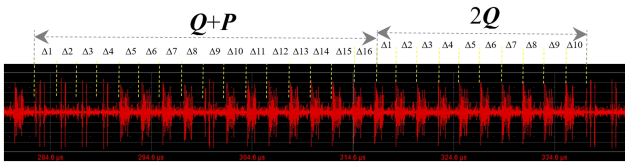

Fig. 3. A part of the *kP* trace measured on the FPGA demonstrating the distinguishability of the EC point addition and the EC point doubling patterns

Fig. 4 shows the first ten atoms of the *Q*+*P* operation from Fig. 3, zoomed in. Each atom starts with a field multiplication, which is the most consuming energy field operation. In our implementation, we used the field multiplier based on the iterative 4-segment Karatsuba multiplication method adapted for the multiplication of prime finite field elements as explained in [16]. Corresponding to this method, both 256 bit long operands are segmented into 4 parts, each 64 bit long. To calculate the field product, nine partial products are calculated and accumulated in the output register of the field multiplier.

Reduction is performed in each clock cycle. If both or even only one of the multiplicands is equal to "0" or "1", the partial products such as 0·0, 0·1, 1·1, or *x*·1 are calculated consuming much less energy than processing 64 bit long multiplicands *x*·*y*. These processes determine the shape of the multiplications and their distinguishability, see Fig. 4.

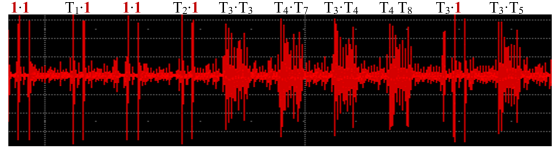

Fig. 4. Atoms Δ1, … , Δ10 of the *Q*+*P* operation from Fig. 3, zoomed in.

Fig. 5 shows the traces measured on the microcontroller. On each trace all PDs and PAs are shown.

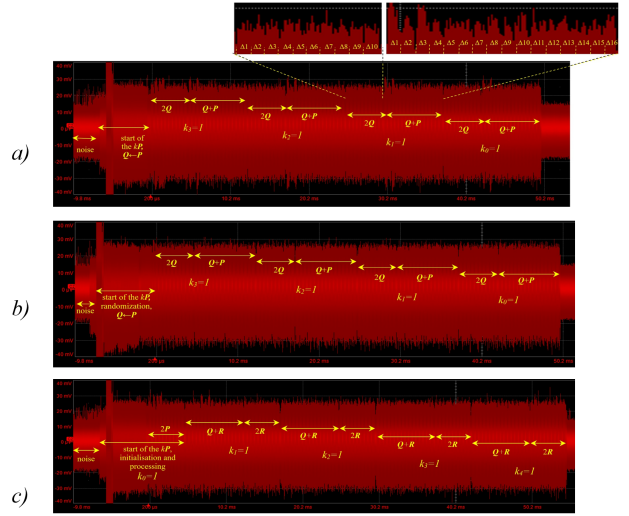

Fig. 5. EM traces measured on the microcontroller for *kP* executions implementing: *a)* – the left-to-right algorithm; *b)* – the left-to-right *kP* algorithm with randomization of projective coordinates; *c)* – the right-to-left algorithm.

In Fig. 5-*a*) the atomic blocks for a PD and a PA are shown also zoomed in, as an example demonstrating their distinguishability in the left-to-right algorithm. It is clearly visible that the atomic patterns Δ1 and Δ3 in *Q*+*P* differ significantly from those of all other atomic blocks. For the right-to-left algorithm (see Fig. 5 - *c*), the first PD and PA operations differ from all other patterns, too. The reasons for the visible distinguishability of atomic blocks are primarily due to operand-dependent energy consumption in field multiplication processing small operand values such as 1 and -3, corresponding to [1]. Only the left-to-right algorithm with randomization of the projective coordinates does not have an easily observable distinguisher (see Fig. 5-*b*), i.e. the shape of all atomic blocks appears visually similar in the EM trace. Thus, the well-known randomisation of elliptic curve point coordinates proposed by [12] as a countermeasure against vertical attacks can also effectively prevent the simple data-bit SCA attacks by replacing the special operand values '1' with long binary numbers.

#### B. Atoms are distinguishable

Additionally, we investigated the distinguishability of atomic block shapes for the left-to-right *kP* algorithm with randomization of projective coordinates (see trace in Fig. 5-*b*) at the beginning of each atom to prove our assumption that the data flow, i.e. the use of different variables for passing data to/from a function, can cause distinguishability of the atoms. This kind of vulnerability is also known as address-bit side-channel leakage phenomena, where the information leakage comes from the addresses of registers or memory locations that are accessed during execution. Regular binary *kP* algorithms, such as the Montgomery ladder[1], are vulnerable to this kind of attacks analysing many *kP* execution traces [22] as well as a single trace [23], at least when implementing *kP* operation in hardware. Atomic patterns [6] are also distinguishable due to the address-bit phenomenon [24]. Different address randomisation methods were proposed in the past [25], [26], [27], [28] and successfully broken by analysing a single trace [29], [30]. Some methods increasing the resistance of hardware implementations to address-bit SCA are proposed in [31].

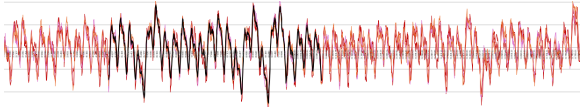

Fig. 6. A part of each atom Δ1, from all four point doubling operations, aligned: traces of the atoms are marked in red; template is marked in black.

The nature of this vulnerability and suitable countermeasures have not investigated yet deeply for embedded processors and microcontrollers. Only a very few works [24], [32], [33] describe such investigations. In our investigation case, although the same operations are performed in each atom, different variables and data-flow paths lead to different address accesses. This can cause observable but very short – only a few clock cycles long – SCA leakage at the start of field operation functions, which allows distinguishing between atoms. Thus, we selected only 24 clock cycles at the beginning of the first atom in the first PD as a *template* for the comparison of all other atoms. Fig. 6 shows a part of each atom Δ1 in each PD (red lines). The traces are aligned. The part of the first atom Δ1, which we selected as template, is marked in black. All atoms Δ1 have a very similar shape at their beginning, i.e. a high correlation with the template is expected. We compared the template with the corresponding parts of other atoms in PDs. A similar shape was observed not for all of them but only for atoms Δ2, Δ4 and Δ7. Fig. 7 shows a part of each atom Δ3 in each PDs (red lines). The traces are aligned and the template is shown as black line (the same template as in Fig. 7). All atoms Δ3 have a very similar shape at their beginning, but it differs from the template in a few clock cycles significantly. Similar differences are observable for other atoms in PDs.

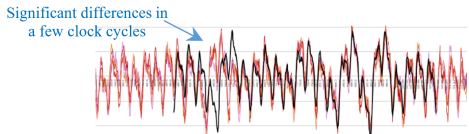

Fig. 7. A part of each atom Δ3 (red lines), from all four PD operations, aligned; template is marked in black.

Based on these observations, we decided to use the template for the calculation of Pearson coefficients for each 24 clock cycles long part of the measured trace to determine all very similar shapes. Thus, we calculated Pearson's coefficients for the whole trace, see Fig. 8-*a)*. The calculated coefficients are mostly in the range of -0.5…0.7, and only a few of them are close to 1.0. We marked the coefficients higher than 0.9 with red dotted lines.

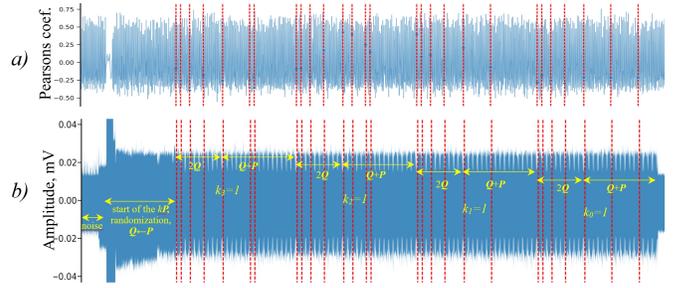

Fig. 8. Distinguishability of PD and PA operations in the left-to-right kP algorithm with randomization of projective coordinates: *a)* - Pearson coefficients calculated for *template* and each 24 clock cycles long part of the measured trace; red dotted lines show the coefficients higher than 0.9; *b)* – labeled measured trace.

The distinguishability of PD and PA operations using the marked correlation coefficients is demonstrated in Fig. 8-*b)*. Each Δ1, Δ2, Δ4, and Δ7 exhibits a high correlation with the template during PD, while in PA, only atom Δ1 maintains this correlation. Conversely, atoms Δ2, Δ4, and Δ7 do not show the same relationship. Additionally, the duration of $Q+P$ is longer than $2Q$. These facts can be successfully exploited to distinguish between the atomic patterns and extract the processed scalar. In our investigations, we focused on a scalar contating all '1', implying that a Q+P operation follows every $2Q$ operation. For other scalars, this is not the case; processing the bit value '0' requires only a $2Q$ operation and processing the bit value '1' requires the sequence of a $2Q$ and a $Q+P$ operations.

## IV. CONCLUSION

In this paper we evaluated the distinguishability of Chevallier-Mames atomic block patterns in binary *kP* algorithms. Both our hardware and software implementations are vulnerable to single-trace SCA attacks. The field multiplication of special small operands is the reason for data-bit vulnerability, which can be easily observed and exploited for key extraction when left-to right algorithm is selected for the implementation. Additionally, we demonstrated that the projective coordinate randomization proposed by Coron [12] can successfully countermeasure this vulnerability but can not prevent the address-bit SCA leakage. We utilize portions of the measured traces related to data transmission in field multiplications as templates to recognize atomic patterns associated with PD and PA operations. Addressing of different registers is the reason for this vulnerability and remains exploitable even when using projective coordinate randomization. Please note that not only field multiplications but also field additions can be potentially such SCA leakage sources, too. Using the proposed here template, the results published in [24], [32], [33] investigating other atomic patterns,

---

[1] It is the notation for *kP* algorithms based on [17]. Please note that the Montgomery ladder is known as resistant to timing and simple SCA attacks [18], [19], [20], assuming that key-dependent addressing of different registers is indistinguishable from SCA point of view [21].

[6], [7], [34] can be improved. Additionally, atomic patterns [35], can be very promising and have to be investigated as well.